\documentclass[12pt,usenames,dvipsnames,a4paper]{article}

\usepackage{soul}

\usepackage[utf8]{inputenc}
\usepackage[T1]{fontenc}

\usepackage{changepage}
\usepackage{amsmath,amsfonts,amssymb}
\usepackage{epsf,amsmath,bbold,amsfonts,stmaryrd}
\usepackage{mathrsfs}
\usepackage{appendix}
\usepackage{caption}
\usepackage{color}
\usepackage{datetime}
\usepackage{float}
\usepackage{graphicx}
\usepackage[colorlinks]{hyperref}
\hypersetup{pageanchor=false,citecolor=red,urlcolor=red}
\usepackage{indentfirst}
\usepackage[numbers,square,comma,sort&compress,merge]{natbib} 
\usepackage{subfig}
\numberwithin{equation}{section}
\usepackage{mathtools}
\usepackage{ytableau}
\usepackage{tabu}
\usepackage{esvect}

\usepackage{calc}

\allowdisplaybreaks

\def\a{\alpha}

\def\b{\beta}
\def\c{\chi}

\def\d{\delta}

\def\e{\varepsilon}

\def\eps{\varepsilon}
\def\f{\frac}
\def\g{\gamma}

\def\G{\Gamma}

\def\l{\left}

\def\mc{\mathcal}

\def\m{\mu}
\def\n{\nu}

\def\p{\partial}

\def\r{\right}
\def\s{\sigma}

\def\x{\xi}

\def\z{\zeta}

\def\be{\begin{equation}}
\def\ee{\end{equation}}

\def\bea{\begin{eqnarray}}
\def\eea{\end{eqnarray}}

\def\ba{\begin{array}}
\def\ea{\end{array}}

\def\bc{\begin{center}}
\def\ec{\end{center}}

\def\bl{\begin{flushleft}}
\def\el{\end{flushleft}}

\def\br{\begin{flushright}}
\def\er{\end{flushright}}

\def\bi{\begin{itemize}}
\def\ei{\end{itemize}}

\def\bt{\begin{tabular}}
\def\et{\end{tabular}}

\makeatletter

\newsavebox\myboxA
\newsavebox\myboxB
\newlength\mylenA

\newcommand*\xoverline[2][0.75]{%
    \sbox{\myboxA}{$\m@th#2$}%
    \setbox\myboxB\null
    \ht\myboxB=\ht\myboxA`
    \dp\myboxB=\dp\myboxA%
    \wd\myboxB=#1\wd\myboxA
    \sbox\myboxB{$\m@th\overline{\copy\myboxB}$}
    \setlength\mylenA{\the\wd\myboxA}
    \addtolength\mylenA{-\the\wd\myboxB}%
    \ifdim\wd\myboxB<\wd\myboxA%
       \rlap{\hskip 0.5\mylenA\usebox\myboxB}{\usebox\myboxA}%
    \else
        \hskip -0.5\mylenA\rlap{\usebox\myboxA}
         {\hskip 0.5\mylenA\usebox\myboxB}%
    \fi}
\makeatother

\def\be{\begin{equation}}
\def\ee{\end{equation}}

\def\bea{\begin{eqnarray}}
\def\eea{\end{eqnarray}}

\def\f{\frac}

\def\p{\partial}

\usepackage{xspace}

\usepackage{xifthen}
\newcommand*\diff{\mathrm{d}} 
\newcommand*\ldiff[2][]{ \ifthenelse{\isempty{#1}}{ \diff
#2}{\diff^#1#2} \,} 
\let\limitint\int 
\renewcommand{\int}{\limitint \!} 

\begin{document}

\begin{titlepage}

\begin{adjustwidth}{-1.3cm}{-.7cm}
\begin{center}
\bf \Large{The particle content of
\texorpdfstring{(scalar curvature)$^2$}~~metric-affine gravity}
\end{center}
\end{adjustwidth}

\begin{center}
\textsc{Georgios K. Karananas}
\end{center}

\begin{center}
\it {Arnold Sommerfeld Center\\
Ludwig-Maximilians-Universit\"at M\"unchen\\
Theresienstra{\ss}e 37, 80333 M\"unchen, Germany}
\end{center}

\begin{center}
\small
\texttt{\small georgios.karananas@physik.uni-muenchen.de} 
\end{center}

\begin{abstract}

Linearizing metric-affine~(scalar curvature)$^2$ gravity---an ``umbrella''
theory that includes as special cases the metrical, Einstein-Cartan, and Weyl
quadratic models---on top of Minkowski spacetime leads to
(numerous) accidental~\emph{gauged}~symmetries. This suggests that the
analysis of the spectrum on flat background is hindered by strong coupling
effects.

Such undesirable symmetries are absent already at the leading nontrivial order
in perturbations on non-flat backgrounds, e.g. de Sitter spacetime, which are
the appropriate ones for studying the particle dynamics of all these
theories. 

\end{abstract}

\end{titlepage}

\section{Introduction}

Metric-affine gravity (MAG) makes the fewest assumptions about the geometry of
spacetime, so in this respect it constitutes the most minimalistic
gravitational formulation. In its full generality, apart from curvature, MAG
is priori also endowed with torsion and nonmetricity. Many more details can
be found in the excellent review~\cite{Hehl:1994ue}.

The focus of this paper is the dynamics of (scalar curvature)$^2$
metric-affine gravity, which encompasses three rather interesting from the
particle physics and cosmology perspective subcases:~\emph{i)}~the metrical
$R^2$ gravity~\cite
{Alvarez-Gaume:2015rwa,Ferreira:2019zzx,Shtanov:2023lci} when both torsion
and nonmetricity vanish;~{\emph{ii)}}~the~(parity-even sector of)
Einstein-Cartan quadratic gravity~\cite{Karananas:2024xja} when the
connection is taken to be metric compatible;~{\emph{iii)}~the Weyl geometric
theory~\cite{Ghilencea:2024usf,Condeescu:2024cjh}~when torsion vanishes and
nonmetricity is purely vectorial.

We show that numerous accidental gauge symmetries emerge when the~
(scalar curvature)$^2$ MAG is linearized on top of Minkowski spacetime. Such
symmetries are not inherited from the parent theory, therefore they are
necessarily broken at higher orders. This constitutes an unsurpassable
obstacle to any consistent interpretation of the flat particle dynamics,
fully resonating with the findings of~\cite
{Casado-Turrion:2023rni,Golovnev:2023zen,Karananas:2024hoh} concerning purely
metrical gravities.~In other words, Minkowski spacetime, although formally a
solution to the field equations, should be completely excluded as a
perturbative background for this type of theories---this is a fact that is
blind to the gravitational formulation.~On the other hand, on backgrounds
with nonvanishing scalar curvature, such as dS spacetime, the theory does not
exhibit emergent gauge redundancies and consequently its particle spectrum,
comprising only the spin-2 massless graviton, can be read-off
unambiguously.

This paper is organized as follows. In Sec.~\ref{sec:dynamics_nonlinear}, we
introduce the full $R^2$ metric-affine gravity and discuss its gauge
symmetries and dynamics. Following more-or-less the corresponding discussion
in~\cite{Glavan:2023cuy}~(see also~\cite{Karananas:2024xja} for similar
considerations in the Einstein-Cartan framework), we demonstrate there that
the theory is equivalent to Einsteinian gravity plus a cosmological constant.
In Sec.~\ref{sec:linear_Minkowski}, we linearize the action on top of
Minkowski spacetime and make explicit the accidental gauge symmetries that
emerge. In Sec.~\ref{sec:dS}, we linearize the action on top of dS where no
emergent gauge redundancies are present. We confirm also at the linearized
level that the action propagates only a massless graviton. We conclude in
Sec.~\ref{sec:conclusions}. Our conventions can be found in Appendix~\ref
{app:Defs_Conventions}.

\section{Metric-affine \texorpdfstring{$R^2$}~~gravity, its gauge
 symmetries and dynamics}
\label{sec:dynamics_nonlinear}

\subsection{The action}

For our purposes here it is most convenient to work in the affine
formulation,\footnote{For the gauge-theoretical formulation of MAG see for
instance~\cite{Percacci:1990wy}.} where the variables describing the
gravitational interaction are the metric $g_{\m\n}$ and connection $\mc
G^\rho_{\ \m\n}$. The former we take with negative signature. The latter is
not assumed to be symmetric in the lower indexes, nor metric-compatible.

The action of the theory under consideration is 
\be
\label{eq:grav_action_full}
S =-\f{1}{f^2}\int \diff^4x\sqrt{g} \mc R^2 \ ,
\ee 
where $f$ is a dimensionless constant, $g=-{\rm det}(g_{\m\n})$, and $\mc R$
is the scalar curvature. All definitions can be found in Appendix~\ref
{app:Defs_Conventions}.

\subsection{Gauge symmetries}

We now discuss the gauge symmetries that~(\ref{eq:grav_action_full}) enjoys,
see e.g.~\cite{Percacci:2020ddy,Glavan:2023cuy}, since this will be important
in the next section where we will linearize on top of Minkowski background.

First, there is invariance under general coordinate transformations, under
which the metric and connection transform in the usual manner, i.e.
\be
\label{eq:diff_trans_full}
g'_{\m\n}(x') = \f{\p x^\kappa}{\p x'^\m}\f{\p x^\lambda}{\p x'^\n}g_{\kappa\lambda}(x) \ ,~~~\mc G'^\rho_{\ \m\n}(x') =\f{\p x'^\rho}{\p x^\s} \f{\p x^\kappa}{\p x'^\m}\f{\p x^\lambda}{\p x'^\n} \mc G^\s_{\ \kappa\lambda}(x)+\f{\p x'^\rho}{\p x^\s}\f{\p^2 x^\s}{\p x'^\m \p x'^\n} \ . 
\ee

Moreover, the scalar curvature---and by association the action---is invariant
under the so-called projective transformations, cf.~\cite
{Iosifidis:2019fsh} and references therein, that only affect the connection
\be
\label{eq:projective_transf_full}
\tilde g_{\m\n}(x) = g_{\m\n}(x) \ ,~~~\tilde{\mc G}^\rho_{\ \m\n} = \mc G^\rho_{\ \m\n}(x) + \d^\rho_\n P_\m (x) \ ,
\ee
with $P_\m$ an arbitrary four-vector.

Finally, the action is invariant under Weyl transformations.\footnote{More on
Weyl transformations in the context of MAG can be found in~\cite
{Iosifidis:2018zwo}.} The metric is rescaled as 
\be
\label{eq:Weyl_transf_full}
\hat g_{\m\n}(x)= e^{2\s(x)}g_{\m\n}(x) \ ,
\ee 
while the connection can either be~\emph{assumed}~to transform inhomogeneously
\be
\label{eq:Weyl_transf_conn_inhom}
\hat{\mc G}^\rho_{\ \m\n}(x) = \mc G^\rho_{\ \m\n}(x) +\d^\rho_{\n}\p_\m \s(x) \ , 
\ee
or~\emph{assumed}~to remain inert
\be 
\label{eq:Weyl_transf_conn_inert}
\hat{\mc G}^\rho_{\ \m\n}(x) = \mc G^\rho_{\ \m\n}(x)  \ .
\ee 
Depending on the behavior of $\mc G$ under Weyl rescalings, it is either
torsion or nonmetricity that transforms inhomogeneously (but not
both).\footnote{I am grateful to Sebastian Zell for discussions on this
point.}

In any event, $\mc R$ transforms covariantly under the Weyl transformations~
(\ref{eq:Weyl_transf_full}) and~(\ref{eq:Weyl_transf_conn_inhom}) or~(\ref
{eq:Weyl_transf_conn_inert})\,\footnote{This is to be contrasted with the
Ricci scalar that transforms inhomogeneously as well known, see e.g.~\cite
{Carroll:2004st}.}
\be
\label{eq:fullR_Weyl}
\hat{\mc{R}} = e^{-2\s}\mc R \ .
\ee 

\subsection{Dynamics of the full nonlinear theory}

The theory~(\ref{eq:grav_action_full}), despite appearances, is
(classically) equivalent to the Einstein-Hilbert action supplemented with a
non-vanishing cosmological constant.\footnote{Surprisingly, the equivalence
between the metric-affine quadratic gravity~(\ref
{eq:grav_action_full}) but~\emph{with a symmetric connection} and the
Einstein-Hilbert action has been known since the late `50s~\cite
{Higgs:1959jua}!}
This can be seen as follows; cf.~Ref.~\cite{Glavan:2023cuy} for a recent nice
discussion along these lines.

First, disentangle the dynamics by introducing an auxiliary field, see
e.g.~\cite
{Tomboulis:1996cy,BeltranJimenez:2019hrm,Pradisi:2022nmh,Glavan:2023cuy,Hinterbichler:2015soa},
$\c$~(with dimension of mass) to recast the gravitational action as
\be
\label{eq:grav_action_auxiliary}
S =-\int \diff^4x \sqrt{g} \l( \c^2 \mc R -\f{f^2\c^4}{4} \r) \ .
\ee 

Then,  take advantage of the Weyl invariance of the
action to set
\be
\c = \f{M_P}{\sqrt 2} \ ,
\ee
so that~(\ref{eq:grav_action_auxiliary}) boils down to its gauge-fixed version 
\be
\label{eq:grav_action_Einstein_2}
S = - \f{M_P^2}{2} \int\diff^4x \sqrt{g} \l( \mc R - \f{f^2 M_P^2}{8} \r) \ .
\ee

Continue by splitting the connection in the standard way as
\be
\label{eq:connection_split}
\mc G^\rho_{\ \m\n} = \Gamma^\rho_{\ \m\n} +\d\G^\rho_{\ \m\n} \ ,
\ee 
where $\G$ are the Christoffel symbols, and $\d\G$ denotes collectively the
torsional and nonmetrical contributions, see the Appendix~\ref
{app:Defs_Conventions} for details. 
Due to~(\ref{eq:connection_split}), the scalar curvature is also decomposed
into the (metrical) Ricci scalar $R$ plus several post-Riemannian pieces
involving torsion ($v_\m,a_\m,\tau_{\m\n\rho}$) and nonmetricity
($u_\m,b_\m,q_{\m\n\rho}$). One finds~\cite
{Langvik:2020nrs,Rigouzzo:2022yan,Rigouzzo:2023sbb}
\begin{align}
\label{eq:curv_decomp}
\mc R &=  R +  \stackrel{\G}{\nabla_\m}\l( 2 v^\m +u^\m - b^\m \r) - \f 2 3 v_\m \l( v^\m + u^\m - b^\m \r) + \f{1}{24} a_\m a^\m  \nonumber \\
 & \qquad+ \f 1 2 \tau_{\m\n\rho}\tau^{\m\n\rho} -\f{11}{72}u_\m u^\m + \f{1}{18} b_\m b^\m + \f 2 9 u_\m b^\m + \f 1 4 q_{\m\n\rho}\l(q^{\m\n\rho} - 2 q^{\rho\m\n}\r) + \tau_{\m\n\rho} q^{\m\n\rho} \ ,
\end{align} 
with $\stackrel{\G}{\nabla^\m}$ the covariant derivative  constructed out of
the Christoffel symbols $\G$.

Plugging the resolution~(\ref{eq:curv_decomp}) of $\mc R$ into~(\ref
{eq:grav_action_Einstein_2}) and dropping full divergences, we obtain
\begin{align}
\label{eq:grav_action_Einstein_3}
S = -\f{M_P^2}{2} &\int\diff^4x \sqrt{g}\Bigg( R - \f{f^2 M_P^2}{8} - \f 2 3 v_\m \l( v^\m + u^\m - b^\m \r) + \f{1}{24} a_\m a^\m + \f 1 2 \tau_{\m\n\rho}\tau^{\m\n\rho} \nonumber \\ 
&\quad\qquad-\f{11}{72}u_\m u^\m + \f{1}{18} b_\m b^\m + \f 2 9 u_\m b^\m + \f 1 4 q_{\m\n\rho}\l(q^{\m\n\rho} - 2 q^{\rho\m\n}\r) + \tau_{\m\n\rho} q^{\m\n\rho}\Bigg) \ .
\end{align} 
Observe that all post-Riemannian pieces appear in the action quadratically and
without derivatives. Their equations of motion can be trivially obtained by
varying the action wrt the corresponding fields $v,a,\tau$
\be
\label{eq:EOM_1}
2v_\m +u_\m -b_\m = 0 \ ,~~a_\m = 0 \ ,~~\tau_{\m\n\rho} + q_{\m\n\rho} = 0 \ ,
\ee
and $u,b,q$
\be
\label{eq:EOM_2}
v_\m+\f{11}{24}u_\m -\f 1 3 b_\m = 0 \ ,~~v_\m +\f 1 3 u_\m +\f 1 6 b_\m = 0 \ ,~~q_{\m\n\rho}-q_{\rho\m\n}-q_{\n\rho\m}+2\tau_{\m\n\rho} = 0 \ .
\ee 
From these we see that $a,\tau$ and $q$ are kinematically projected
to zero
\be
\label{eq:ataut}
a_\m = 0 \ ,~~~\tau_{\m\n\rho} = 0 \ ,~~~q_{\m\n\rho} = 0 \ ,
\ee 
whereas\,\footnote{\label{foot:projective}Notice that the equations of motion
for the vectors $v_\m,u_\m,b_\m$, see~(\ref{eq:EOM_1},\ref{eq:EOM_2}), as
well as the solutions~(\ref{eq:ub}) are manifestly invariant under the
projective transformation~(\ref{eq:projective_transf_full}), that in terms of
these fields reads~\cite{Rigouzzo:2022yan}
\be
\label{eq:proj_sym}
 \tilde v_\m =v_\m +3 P_\m \ ,~~~\tilde u_\m = u_\m - 8P_\m \ ,~~~\tilde b_\m = b_\m - 2P_\m \ .
\ee
On the other hand, the rest of the components of the connection do not transform
\be
\tilde a_\m = a_\m \ ,~~~\tilde \tau_{\m\n\rho} = \tau_{\m\n\rho} \ ,~~~\tilde q_{\m\n\rho} = q_{\m\n\rho} \ . 
\ee
}
\be
\label{eq:ub}
u_\m = -\f 8 3 v_\m \ ,~~~b_\m = -\f 2 3 v_\m \ ,
\ee 
meaning that 
\be
\label{eq:ub_relation}
u_\m = 4 b_\m \ .
\ee

Plugging~(\ref{eq:ataut},\ref{eq:ub}) in~(\ref{eq:grav_action_Einstein_3}), we
find that torsion and nonmetricity completely disappear from the action: 
\be
\label{eq:grav_action_Einstein_4}
S = - \f{M_P^2}{2} \int \diff^4 x \sqrt{g}\l(  R - \f{f^2 M_P^2}{8} \r) \ .
\ee 
This is exactly metrical General Relativity (GR), 
 therefore, the theory propagates~\emph{the massless
graviton, only}~\cite{Glavan:2023cuy}.

Some comments are in order. First, had we started with curvature-squared
gravity in the metric formalism, we would have gotten an additional
propagating massless scalar~\cite{Alvarez-Gaume:2015rwa}, associated with the
conformal mode of the metric. For nonvanishing torsion or nonmetricity, the
curvature is a Weyl-covariant object, see (\ref{eq:fullR_Weyl}), and
consequently this extra field of gravitational origin is absent. Second,  as
obvious from~(\ref{eq:grav_action_full}), the theory formally admits as
solution a flat metric and vanishing connection, i.e. Minkowski spacetime. As
we shall show in the next section, linearizing the action on top of it
results into accidental gauge symmetries and a seemingly trivial from the
perspective of particle dynamics theory. Third, we find it remarkable that
the ``physical branch'' of the theory with nonvanishing curvature is
automatically singled out by utilizing an auxiliary field. However, in order
to avoid any misconceptions and confusions, we shall not make use of this
method again in the remainder of the paper. Actually, we could have avoided
it altogether; nevertheless, we found it the most clean and illuminating.

\section{Linearizing on top of Minkowski spacetime: accidental
 gauge symmetries }
 \label{sec:linear_Minkowski}

There are (at least) two straightforward ways to see why perturbing the theory
on top of Minkowski is a no-go.

\textbf{The first route}~is to start from the action~(\ref
 {eq:grav_action_full}) and plug in~(\ref{eq:curv_decomp}), which yields
\begin{align}
\label{eq:grav_action_Einstein_3}
S = -&\f{1}{f^2} \int\diff^4x \sqrt{g}\Bigg( R +  \stackrel{\G}{\nabla_\m}\l( 2 v^\m +u^\m - b^\m \r) - \f 2 3 v_\m \l( v^\m + u^\m - b^\m \r) + \f{1}{24} a_\m a^\m  \nonumber \\ 
& + \f 1 2 \tau_{\m\n\rho}\tau^{\m\n\rho}-\f{11}{72}u_\m u^\m + \f{1}{18} b_\m b^\m + \f 2 9 u_\m b^\m + \f 1 4 q_{\m\n\rho}\l(q^{\m\n\rho} - 2 q^{\rho\m\n}\r) + \tau_{\m\n\rho} q^{\m\n\rho}\Bigg)^2 \ .
\end{align} 

We now consider excitations of the fields on top of the flat background
\be
\label{eq:Minkowski_background}
g_{\m\n} = \eta_{\m\n} \ ,~~~\mc G^\rho_{\ \m\n} = 0 \ ,
\ee 
with $\eta_{\m\n}={\rm diag}(1,-1,-1,-1)$ the Minkowski metric.\footnote
{Owing to~(\ref{eq:connection_split}), the fact that on the background the
full connection is zero of course means that 
\be
v_\m=a_\m=\tau_{\m\n\rho}=u_\m=b_\m=q_{\m\n\rho} = 0 \ .
\ee}

The action~(\ref{eq:grav_action_Einstein_3}) to quadratic order in the
fluctuations reads
\begin{align}
\label{eq:quad_act_Mink}
S_2 = -\f{1}{f^2} \int \diff^4x \Big[ &\l(\p_\m\p_\n h^{\m\n} - \square h\r)\l(\p_\rho\p_\s h^{\rho\s} - \square h\r) \nonumber\\
&\qquad+2 \l(\p_\m\p_\n h^{\m\n} - \square h\r) \p_\rho \l(2 v^\rho+u^\rho-b^\rho\r)\nonumber\\
&\qquad\qquad+ \p_\m\l(2 v^\m+u^\m-b^\m\r)\p_\n\l(2 v^\n+u^\n-b^\n\r)  \Big] \ ,
\end{align} 
where $h_{\m\n}$ is the metric perturbation, $h=h^\m_\m$ its trace and
$\square =\p^\m \p_\m$; indexes are raised and lowered with the Minkowski
metric. In an abuse of notation we retained the same symbols $v_\m$ and
$u_\m,b_\m$ for the excitations of torsion and non-metricity, respectively.

As expected, the symmetries of the parent theory~(\ref
{eq:grav_action_full}) have been passed down (in their linearized form) to
$S_2$, meaning that the latter is invariant under: 
\begin{itemize}
\item[\emph{(i)}]~diffeomorphisms, that act on $h_{\m\n}$ and $V_\m = v_\m,u_\m,b_\m$, as 
\be
\d h_{\m\n} = \p_\m \x_\n +\p_\n \x_\m \ ,~~~\d V_\m = \x^\n \p_\n V_\m + V_\n \p_\m \x^\n \ ,
\ee
with $\x^\m$ a four-vector; 

\item[\emph{(ii)}]~projective transformations that act on the various fields
 as (see also footnote~\ref{foot:projective})
\be
\label{eq:projective_linear}
\d_P h_{\m\n} = 0 \ ,~~~\d_P v_\m = 3 P_\m \ ,~~~\d_P u^\m = -8 P_\m \ ,~~~\d_P b_\m = -2 P_\m \ ;
\ee

\item[\emph{(iii)}]~Weyl rescalings, with the metric perturbation transforming as
\be
\d_{\rm W} h_{\m\n} = 2\s \eta_{\m\n} \ ,
\ee 
and depending on how the connection is assumed to behave, see~(\ref
{eq:Weyl_transf_conn_inhom},\ref{eq:Weyl_transf_conn_inert}),  it is either
the torsion vector that transforms inhomogeneously and the nonmetricity
vectors remain intact 
\be
\label{eq:weyl_tors}
\d_W v_\m = 3 \p_\m \s \ ,~~~\d_W u_\m = 0 \ ,~~~\d_W b_\m = 0 \ ,
\ee
or the other way around
\be
\label{eq:weyl_nonmet}
\d_W v_\m = 0 \ ,~~~\d_W u_\m = 8\p_\m\s \ ,~~~\d_W b_\m = 2\p_\m \s \ .
\ee

\end{itemize}

This is not the whole story though. In addition to~\emph{(i)-(iii)}, the
quadratic action~(\ref{eq:quad_act_Mink}) exhibits a number of accidental
gauge redundancies. Obviously, even one emergent invariance---just like it
happens in metrical $R^2$ gravity~\cite{Karananas:2024hoh}---is alarming. For
the sake of completeness we discuss in details the situation, simply because
it is more profound in the full MAG $R^2$ gravity than in its metrical
subclass.

First of all, $h_{\m\n}$ enters $S_2$ only through the transverse operator
$\p_\m\p_\n-\eta_{\m\n}\square$. This results into an accidental tensorial
gauge invariance of~(\ref{eq:quad_act_Mink}) under the following shift of the
graviton~\cite{Karananas:2024hoh}
\be
\d_{\rm t} h_{\m\n} = \z^{TT}_{\m\n} \ ,~~~{\rm with}~~~\p^\m \z^{TT}_{\m\n} = 0 \ ,~~~\eta^{\m\n}\z^{TT}_{\m\n} = 0 \ .
\ee

Second, the mass terms for the torsion and nonmetricity vectors
$v_\m,u_\m,b_\m$ are absent from~(\ref{eq:quad_act_Mink}). This gives rise to
yet another, vectorial, accidental gauge symmetry. Namely, the quadratic
action is invariant under 
\be
\label{eq:accidental_vectorial}
\d_Q h_{\m\n} = 0 \ ,~~~\d_Q v_\m =c_v Q_\m \ ,~~~\d_Q u_\m = c_u Q_\m\ ,~~~\d_Q b_\m = c_b Q_\m \ , 
\ee 
with the constants $c_v,c_u$ and $c_b$ subject to 
\be
2c_v+c_u-c_b = 0 \ , 
\ee
and
$Q_\m$ a four-vector.\footnote{Interestingly, for the specific choice 
\be
c_v=1 \ ,~~~c_u = 0\ ,~~~c_b= 2 \ ,
\ee
 the accidental vectorial transformation~(\ref{eq:accidental_vectorial}) combined with the projective one~(\ref{eq:projective_linear}), define the ``extended projective transformation''
\be
\d_{\rm EP} h_{\m\n} = 0  \ ,~~~\d_{\rm EP} v_\m = 3P_\m + Q_\m \ ,~~~\d_{\rm EP} u_\m = -8P_\m \ ,~~~\d_{\rm EP}b_\m = -2P_\m +2 Q_\m \ ,
\ee
introduced in~\cite{Barker:2024dhb}.}

Finally, and this is very important, notice that the axial torsion $a_\m$, as
well as the tensors $\tau_{\m\n\rho}$ and $q_{\m\n\rho}$ have completely
disappeared from~(\ref{eq:quad_act_Mink}); these fields vanish in the full
theory by virtue of their equations of motion~(\ref{eq:ataut}), as we showed
in the previous section. Their absence from the quadratic action brings about
more accidental gauge redundancies---associated with the spin-1, spin-2 and
spin-3 sectors of the theory.

It turns out the particle spectrum of at the quadratic level is empty on
Minkowski spacetime,\footnote{See~\cite
{Hell:2023mph,Golovnev:2023zen,Karananas:2024hoh} for the purely metrical
situation.} something that can be immediately seen from the equations of
motion that follow from~(\ref{eq:quad_act_Mink}). Moreover, expanding the
action to higher orders in perturbations one sees that all accidental
symmetries are explicitly violated, a clean-cut signal of strong coupling.

\textbf{The second, faster, route}~to the exact same conclusion requires the
 bare minimum amount of calculations and utilizes spin-projection operators
 for the full connection $\mc G$. 
 
Evaluating~(\ref{eq:grav_action_full}) on top of~(\ref
{eq:Minkowski_background}), we find a remarkably simple expression
\be
\label{eq:lin_Mink_Gamma}
S_2 = -\f{1}{f^2} \int\diff^4x \Big[ \l(\p_\m \mc G^{\m\n}_{~~\n} -\p^\m \mc G^\n_{~\n\m}\r)\l(\p_\rho \mc G^{\rho\s}_{~~\s} -\p^\rho \mc G^\s_{~\s\rho}\r)\Big] \ .
\ee 
Note that, had we intended to actually study the spectrum, we should have
introduced a source for the connection.

We proceed by projecting-out the spin-J component(s) of $\mc G$ with the use
of the standard orthonormal operators, denoted hereafter by $P^J_a$. Being a
3-index object without symmetries, the connection carries 64 degrees of
freedom in 4 spacetime dimensions. These correspond~\cite
{Percacci:2020ddy} to one spin-3 field~(7 components), five spin-2 fields~
(25 components), nine spin-1 fields~(27 components) and five spin-0 fields~
(5 components):
\be
\mc G_{\m\n\rho} = \l(P^3 + \underbrace{P_1^2 + \ldots + P_5^2}_{\rm 5~operators} + \underbrace{P_1^1+\ldots +P_9^1}_{\rm 9~operators} + \underbrace{P_1^0 +\ldots+P_5^0}_{\rm 5~operators} \r)_{\m\n\rho\a\b\g} \mc G^{\a\b\g} \ .
\ee 
In the above, we have heavily simplified and condensed the notation. 

Now, it suffices to only consider the spin-3 component of the
connection:\footnote{This is associated with the $q_{\m\n\rho}$ nonmetrical
tensor.} this must vanish on-shell, so it must enter the linearized action
quadratically and moreover, without derivatives. If it does not appear at
all, then owing to the orthogonality of the projectors, Eq.~(\ref
{eq:lin_Mink_Gamma}) exhibits invariance under 
\be
\d \mc G_{\m\n\rho} = P^3_{\m\n\rho\a\b\g}\x^{\a\b\g} \ ,
\ee 
with $\x_{\a\b\g}$ arbitrary. This accidental redundancy---completely
unrelated to diffs, projective or Weyl symmetries---imposes irrelevant
constraints on the source that we should have included if we were to study
the spectrum. Since the coupling is universal, these~\emph{affect all spin
subsectors,} which in turn renders the analysis inconclusive. 

Indeed, using the explicit form~\cite{Percacci:2020ddy} of the
projectors,\footnote{See also the ancilliary Mathematica file of~\cite
{Percacci:2020ddy} on arXiv: \url{http://arxiv.org/abs/1912.01023}.} it is a
straightforward exercise to verify the complete absence of a spin-3 sector,
which demonstrates the breakdown of the perturbative analysis of the spectrum
on Minkowski spacetime.

\section{Linearizing on top of de Sitter spacetime}
\label{sec:dS}

The situation on ``healthy'' backgrounds changes radically and we show that
concretely by perturbing the action on top of dS. 

The logic is more or less the same as in the previous section. Start from
(\ref{eq:grav_action_full}), decompose the scalar curvature as in~(\ref
{eq:curv_decomp}) and instead of considering perturbations on top of
Minkowski~(\ref{eq:Minkowski_background}), take as background dS spacetime
\be
g_{\m\n} = \bar g_{\m\n} \ ,~~~\mc G^\rho_{\m\n} = \bar \G^\rho_{\m\n} \ ,
\ee 
with $\bar \G^\rho_{\ \m\n}$ the Christoffel symbols evaluated on the
background metric. Indexes in what follows are raised and lowered with $\bar
g_{\m\n}$ and its inverse $\bar g^{\m\n}$.

The quadratic action breaks naturally into three parts
\be
\label{eq:S2_dS_linearized}
\bar S_2 = \bar S_2^1 + \bar S_2^2 +\bar S_2^3 \ .
\ee
The first depends only on the metric perturbation
\begin{align}
\label{eq:EH_dS_linearized}
\bar S_2^1 = \int \diff^4 x \sqrt{\bar g}&\Bigg [ \f{\bar R}{2}\big(\mathcal D_\rho h_{\m\n}\mathcal D^\rho h^{\m\n}-2\mathcal D^\m h_
  {\m\n}\mathcal D_\rho h^{\n\rho} +\mathcal D^\m h \mathcal D^\n h_
  {\m\n}\big) \nonumber \\
&+\f{\bar R^2}{12}\l( h_{\m\n}h^{\m\n}-\f{h^2}{4} \r) - \l(\mathcal D_\m\mathcal D_\n h^
   {\m\n}-\mathcal D^2 h\r)\l(\mathcal D_\rho\mathcal D_\s h^
   {\rho\s}-\mathcal D^2 h\r) \Bigg] \ ,
\end{align}
the second contains the mixings between $h_{\m\n}$ and $v_\m,u_\m,b_\m$
\begin{align}
\label{eq:EH_dS_linearized}
\bar S_2^2 = - &\int \diff^4 x \sqrt{\bar g} \Big [ 2 \l(\mathcal D_\m\mathcal D_\n h^
   {\m\n}-\mathcal D^2 h\r)\mathcal D_\rho\l(2 v^\rho+u^\rho-b^\rho\r)\\
   &\qquad\qquad\qquad\qquad\qquad\qquad\qquad\qquad\quad-\f{\bar R}{2} h \mathcal D_\m\l(2 v^\m+u^\m-b^\m\r)  \Big] \ , \nonumber
\end{align}
and the last comprises only post-Riemannian contributions  
\begin{align}
\bar S_2^3 = -\int \diff^4 x\sqrt{\bar g}\Bigg[
& \mathcal D_\m\l(2 v^\m+u^\m-b^\m\r)\mathcal D_\n\l(2 v^\n+u^\n-b^\n\r)  \nonumber\\
&- \bar R\l( \f 4 3 v_\m \l( v^\m + u^\m - b^\m \r)-\f{1}{12}a_\m a^\m -\tau_{\m\n\rho}\tau^{\m\n\rho} +\f{11}{36}u_\m u^\m  \r. \nonumber \\
&\qquad\l.- \f{1}{9} b_\m b^\m - \f 4 9 u_\m b^\m - \f 1 2 q_{\m\n\rho}\l(q^{\m\n\rho} - 2 q^{\rho\m\n}\r) - 2\tau_{\m\n\rho} q^{\m\n\rho}\r)\Bigg]   \ . 
\end{align} 
In a self-explanatory notation, $\bar g$ is (minus) the determinant of the dS
metric, $\mathcal D_\m$ the corresponding covariant derivative involving
$\bar \G$, $\mathcal D^2=\bar g^{\m\n}\mathcal D_\m\mathcal D_\n$, $h=\bar g^
{\m\n}h_{\m\n}$, while $\bar R=12\Lambda$ is the background curvature. We
have trivially rescaled all fields in order to absorb the overall constant
$f$.~\emph{In all the considerations that follow we assume that $\bar R\neq
0$.}

Notice that in the quadratic action the metric perturbation appears as it
should, all post-Riemannian fields are present, and so are the mass terms for
the vectors. It can be verified that Eq.~(\ref{eq:S2_dS_linearized}) is
invariant under linearized diffeomorphisms, projective transformations and
Weyl rescalings, only. No accidental gauge symmetries emerge on dS, in
complete analogy with metric $R^2$ gravity: the background curvature is a
``regulator''~\cite{Alvarez-Gaume:2015rwa} that as long as it does not
vanish, degrees of freedom are not evanescent and the theory cannot reach the
strong-coupling point. These observations constitute an important sanity
check for the whole consistency of the determination of the particle
spectrum.  

Having ensured that dS is an admissible background, we turn to elucidating the
dynamics of the linearized theory and demonstrate that it propagates only the
massless graviton, as expected from the considerations of Sec.~\ref
{sec:dynamics_nonlinear} and the full-blown Hamiltonian analysis of~\cite
{Glavan:2023cuy}.

We found it simpler and ``safer''\,\footnote{Since we are not utilizing the
auxiliary field method, nor are we gauge-fixing the Weyl invariance, the
equations of motion are not purely algebraic.} to work with the equations of
motion, but, of course, the exact same results can be easily obtained by
working with the action.\footnote{In principle, one may even work with dS
spin-projection operators. These were constructed in~\cite
{Kuzenko:2019kqw,Hutchings:2024qqf}.}~Varying~(\ref{eq:S2_dS_linearized}) wrt
$a,\tau,q$ results into the following equations of motion, that coincide with
the corresponding ones found in Sec.~\ref{sec:dynamics_nonlinear}, see~(\ref
{eq:EOM_1},\ref{eq:EOM_2}),
\be
a_\m = 0 \ ,~~~ \tau_{\m\n\rho} + q_{\m\n\rho} = 0 \ ,~~~q_{\m\n\rho}-q_{\rho\m\n}-q_{\n\rho\m}+2\tau_{\m\n\rho} = 0 \ ,
\ee 
meaning that these fields vanish dynamically.

The equations of motion for $h_{\m\n},v_\m,u_\m$ and $b_\m$ evaluated on the
above are
\begin{align}
\label{eq:dS_eom_h}
\bar R\l[\mathcal D^2 h_{\m\n} -\mathcal D_\m \mathcal D_\rho h^\rho_\n -\mathcal D_\n \mathcal D_\rho h^\rho_\m + \f 1 2 \mathcal D_\m \mathcal D_\n h + \f{1}{2}\bar g_{\m\n} \mathcal D_\rho \mathcal D_\s h^{\rho\s} - \f{\bar R}{6} \l(h_{\m\n} -\f{\bar g_{\m\n}}{4} h \r) \r]& \nonumber \\
+2\l(\mathcal D_\m \mathcal D_\n -\bar g_{\m\n}\mathcal D^2\r)\l(\mathcal D_\rho \mathcal D_\s h^{\rho\s} - \mathcal D^2 h\r) +\l(\mathcal D_\m \mathcal D_\n -\bar g_{\m\n}\mathcal D^2\r) \mathcal D_\rho\l( 2v^\rho +u^\rho -b^\rho \r)& 
\nonumber \\
 -\bar g_{\m\n}\f{\bar R}{2} \mathcal D_\rho \l(2 v^\rho +u^\rho - b^\rho \r) =  0  \ ,&\\
 \nonumber\\
 \label{eq:dS_eom_v}
\p_\rho \l( \mathcal D_\m \l(2 v^\m +u^\m - b^\m \r) + \mathcal D_\m \mathcal D_\n h^{\m\n} - \mathcal D^2 h -\f{\bar R}{4} h \r) +\f{\bar R}{3} \l(2v_\rho + u_\rho - b_\rho \r)  = 0 \ ,& \\  
\nonumber\\
\label{eq:dS_eom_u}
\p_\rho \l( \mathcal D_\m \l(2 v^\m +u^\m - b^\m \r) + \mathcal D_\m \mathcal D_\n h^{\m\n} - \mathcal D^2 h -\f{\bar R}{4} h \r) +\f{\bar R}{3} \l(2v_\rho + \f{11}{12} u_\rho -\f 2 3 b_\rho \r)  = 0 \ ,&\\
\nonumber\\
\label{eq:dS_eom_b}
\p_\rho \l( \mathcal D_\m \l(2 v^\m +u^\m - b^\m \r) + \mathcal D_\m \mathcal D_\n h^{\m\n} - \mathcal D^2 h -\f{\bar R}{4} h \r) +\f{\bar R}{3} \l(2v_\rho + \f{2}{3} u_\rho + \f 1 3 b_\rho \r)  = 0 \ ,&
\end{align}
respectively. 
  
Subtracting~(\ref{eq:dS_eom_b}) from~(\ref{eq:dS_eom_u}), we find
\be
\label{eq:dS_ub_relation}
u_\m = 4 b_\m \ ,
\ee 
which, to no surprise, is the relation~(\ref{eq:ub_relation}) obtained from
the equations of motion of the full nonlinear theory. Adding~(\ref
{eq:dS_eom_u}) and~(\ref{eq:dS_eom_b}) and using the above does not bring in
any new information, since it boils down to~(\ref{eq:dS_eom_v}) upon
plugging-in~(\ref{eq:dS_ub_relation}):  
\be
\label{eq:dS_eom_v_2}
\p_\rho \l( \mathcal D_\m \l(2 v^\m +3b^\m \r) + \mathcal D_\m \mathcal D_\n h^{\m\n} - \mathcal D^2 h -\f{\bar R}{4} h \r) +\f{\bar R}{3} \l(2v_\rho + 3b_\rho \r)  = 0 \ .
\ee 
Acting on this expression with $\mathcal D_\s$ and antisymmetrizing in $\s$
and $\rho$, one finds that 
\be
\mathcal D_\rho (2v_\s + 3b_\s) = \mathcal D_\s (2v_\rho + 3b_\rho) \ ,
\ee
implying that 
\be
\label{eq:dS_bv_relation}
b_\m = -\f 2 3 \l( v_\m -3\p_\m \varphi \r) \ ,
\ee 
with $\varphi$ a scalar with mass-dimension one; the relative coefficient
between $v_\m$ and $\p_\m\varphi$ was chosen such that the expression
conforms  with the Weyl transformations~(\ref{eq:weyl_tors},\ref
{eq:weyl_nonmet}).\footnote{Note that~(\ref{eq:dS_bv_relation}) is consistent
with~(\ref{eq:ub}), the former boiling down to the latter for
$\varphi=$~constant.} 

Substituting the solution~(\ref{eq:dS_bv_relation}) into~(\ref
{eq:dS_eom_v_2}), we obtain
\be
\label{eq:dS_eom_v_3}
\p_\rho \l(6\mathcal D^\m \p_\m \varphi + \mathcal D_\m \mathcal D_\n h^{\m\n} - \mathcal D^2 h -\f{\bar R}{4} h \r) + 2\bar R \p_\rho \varphi = 0 \ ,
\ee 
while the equations of motion for the metric perturbation~(\ref
{eq:dS_eom_h}) on~(\ref{eq:dS_ub_relation},\ref{eq:dS_bv_relation}) become 
\begin{align}
\label{eq:dS_eom_h_2}
\bar R\l(\mathcal D^2 h_{\m\n} -\mathcal D_\m \mathcal D_\rho h^\rho_\n -\mathcal D_\n \mathcal D_\rho h^\rho_\m + \f 1 2 \mathcal D_\m \mathcal D_\n h + \f{1}{2}\bar g_{\m\n} \mathcal D_\rho \mathcal D_\s h^{\rho\s} - \f{\bar R}{6} \l(h_{\m\n} -\f{\bar g_{\m\n}}{4} h \r) \r)& \nonumber \\
+2\l(\mathcal D_\m \mathcal D_\n -\bar g_{\m\n}\mathcal D^2\r)\l(\mathcal D_\rho \mathcal D_\s h^{\rho\s} - \mathcal D^2 h\r) +6\l(\mathcal D_\m \mathcal D_\n -\bar g_{\m\n}\mathcal D^2\r) \mathcal D^\rho \p_\rho \varphi & \nonumber \\
-3 \bar g_{\m\n}\bar R \mathcal D^\rho \p_\rho\varphi =0  &\ .  
\end{align}
It is important to remember that there is still residual Weyl invariance,
meaning that $\varphi$ is redundant. This can be readily seen by performing
the following change of variables\,\footnote{This is of course equivalent to
setting $\varphi=$constant.}
\be
\label{eq:dS_lin_Weyl}
h_{\m\n} = \hat h_{\m\n} + 2 \bar g_{\m\n} \varphi \ ,
\ee
that completely eliminates the scalar from~(\ref{eq:dS_eom_v_3},\ref
{eq:dS_eom_h_2})
\begin{align}
\label{eq:dS_constr_h}
\p_\rho \l(\mathcal D_\m \mathcal D_\n \hat h^{\m\n} - \mathcal D^2 \hat h -\f{\bar R}{4} \hat h \r) = 0 \ ,& \\
\bar R\l(\mathcal D^2 \hat h_{\m\n} -\mathcal D_\m \mathcal D_\rho \hat h^\rho_\n -\mathcal D_\n \mathcal D_\rho \hat h^\rho_\m + \f 1 2 \mathcal D_\m \mathcal D_\n \hat h + \f{1}{2}\bar g_{\m\n} \mathcal D_\rho \mathcal D_\s \hat h^{\rho\s} - \f{\bar R}{6} \l(\hat h_{\m\n} -\f{\bar g_{\m\n}}{4} \hat h \r) \r)& \nonumber \\
+2\l(\mathcal D_\m \mathcal D_\n -\bar g_{\m\n}\mathcal D^2\r)\l(\mathcal D_\rho \mathcal D_\s \hat h^{\rho\s} - \mathcal D^2 \hat h\r) =0 \ ,&
\end{align}
which, when combined appropriately, yield 
\be
\mathcal D^2 \hat h_{\m\n} -\mathcal D_\m \mathcal D_\rho \hat h^\rho_\n -\mathcal D_\n \mathcal D_\rho \hat h^\rho_\m + \mathcal D_\m \mathcal D_\n \hat h + \bar g_{\m\n} \l(\mathcal D_\rho \mathcal D_\s \hat h^{\rho\s} - \mathcal D^2 \hat h \r)- \f{\bar R}{6} \l(\hat h_{\m\n} +\f{\bar g_{\m\n}}{2} \hat h \r)  = 0 \ .
\ee 
These are precisely the linearized equations for the graviton on top of a dS
spacetime, as follow from the Einstein-Hilbert action.

\section{Conclusions} 
\label{sec:conclusions}

In this paper we discussed aspects of the metric-affine $R^2$ gravity. First,
we gave an overview of the nonlinear symmetries and dynamics of the theory.
As well-known, the full action is diffeomorphism-, Weyl- and projective-
invariant, and captures the dynamics of a massless graviton.  

Then, we showed that, perturbatively, on top of Minkowski spacetime the action
exhibits a number of accidental gauge redundancies, translating into
inconclusive conclusions concerning the propagating modes. 

Finally, we carried out the linearization exercise on top of de Sitter
spacetime and showed that, as expected, no accidental symmetries emerge.
Thus, there is no obstacle to determining the particle spectrum, which was
explicitly shown to contain the massless graviton.

\section*{Acknowledgements}

I am grateful to Will Barker, Dra\v{z}en Glavan, Misha Shaposhnikov and
Sebastian Zell for correspondence, discussions and important comments on the
manuscript.

\appendices

\section{Definitions and Conventions}
\label{app:Defs_Conventions}

\begin{itemize}

\item Apart from the signature of the metric that we take to be mostly minus,
 we use the conventions of~\cite{Rigouzzo:2022yan,Rigouzzo:2023sbb}.

\item Round (square) brackets denote normalized symmetrization
 (antisymmetrization) of the enclosed indexes.

\item The covariant derivative of a contravariant vector is defined as 
\be
\nabla_\m V^\n = \p_\m V^\n + \mc G^\n_{\ \m \rho} V^\rho \ ,
\ee
where $\mc G^\rho_{\ \m\n}$ is the affine connection. 

\item The affine curvature tensor reads
\be
\label{eq:curv_tens_def}
\mc R^\rho_{~\sigma\m\n} = \p_\m \mc G^\rho_{~\n\s} - \p_\n \mc G^\rho_{~\m\sigma} + \mc G^\rho_{~\m\lambda} \mc G^\lambda_{~\n\sigma} -\mc G^\rho_{~\n\lambda} \mc G^\lambda_{~\m\sigma} \ ,
\ee 
and the scalar curvature is given by its trace 
\be
\label{eq:R_def}
\mc R = g^{\sigma\n} \d^\m_\rho \mc R^\rho_{~\sigma\m\n} \ . 
\ee 

\item Torsion is the antisymmetric part of the full connection
\be
T^\rho_{\ \m\n} = \mc G^\rho_{\ \m\n} - \mc G^\rho_{\ \n\m} \ .
\ee
To simplify computations, it is useful to decompose the above as
\be
T_{\m\n\rho} = \f 1 6 E_{\m\n\rho\s} a^\s -\f 2 3 g_{\m[\n}v_{\rho]} +\tau_{\m\n\rho} \ ,
\ee 
where 
\begin{align}
\label{eq:torsion_components}
&a^\m = E^{\m\n\rho\sigma}T_{\n\rho\sigma} \ ,~~v_\m = T_{\n\m}^{~~\n} \ ,\\
&\tau_{\m\n\rho} = \f 2 3 \l( T_{\m\n\rho} -v_{[\n}g_{\rho]\m} -T_{[\n\rho]\m} \r) \ ,~~~{\rm with}~~~\tau^\m_{\ \n\m} = E^{\m\n\rho\s}\tau_{\n\rho\s} = 0 \ ,
\end{align}
are the axial vector, trace vector and reduced torsion tensor,
respectively; we also introduced 
\be
E^{\m\n\rho\s} = \f{\eps^{\m\n\rho\s}}{\sqrt{g}} \ ,~~~E_{\m\n\rho\s} =\sqrt{g} \eps_{\m\n\rho\s} \ ,
\ee
with $\e$ the totally antisymmetric symbol. 

\item Nonmetricity is defined as the covariant derivative of the metric
\be
Q_{\m\n\rho} = \nabla_{\m}g_{\n\rho} \ ,
\ee
which can be expressed as
\be
Q_{\m\n\rho} = \f{1}{18}\l(g_{\n\rho} \l(5 u_\m - 2 b_\m \r) + 2 g_{\m(\n}\l(4 b_{\rho)} - u_{\rho)} \r) \r) + q_{\m\n\rho} \ ,
\ee
where 
\begin{align}
\label{eq:nonmetricity_components}
&b_\m = Q_{\n\m}^{~~\n} \ ,~~u_\m = Q_{\m\n}^{~~\n} \ ,\\
&q_{\m\n\rho} = Q_{\m\n\rho} - \f{1}{18}\l(g_{\n\rho}(5u_\m-2b_\m) + 2g_{\m(\n}(4b_{\rho)} - u_{\rho)}) \r) \ ,~~~{\rm with}~~~q^\m_{\ \m\n} = q^{\m}_{\ \n \m} = 0 \ ,
\end{align} 

are the nonmetrical vectors and traceless tensor. 

\item The full affine connection can be split into Riemannian plus
 post-Riemannian pieces
\be
\label{eq:app_conn_decomp}
\mc G^\rho_{\ \m\n} = \G^\rho_{~\m\n} + \d\G^\rho_{\ \m\n} \ ,
\ee
where
\be
\label{eq:christoffel}
\G^\rho_{~\m\n} = \f 1 2 g^{\rho\sigma}\l(\p_\m g_{\n\sigma}+\p_\n g_{\m\sigma}-\p_\sigma g_{\m\n} \r) \ ,
\ee
are the usual Christoffel symbols, and 
\be
\label{eq:post_riem}
\d\G^\rho_{\ \m\n} = K^\rho_{\ \m\n} + J^\rho_{\ \m\n} \ ,
\ee
with $K$ and $J$ the so-called controrsion and disformation tensors. The
former is associated with torsion and in terms of~(\ref
{eq:torsion_components}) reads as
\be
\label{eq:contorsion}
K_{\m\n\rho} = \f{1}{12} E_{\m\n\rho\sigma}a^\sigma - \f 2 3 g_{\n[\m}v_{\rho]} +2 \tau_{[\m|\n|\rho]} \ ,
\ee
while the latter is associated with nonmetricity and in terms of~(\ref
{eq:nonmetricity_components}) is given by
\begin{align}
\label{eq:disformation}
&J_{\m\n\rho} =  \f 1 9 g_{\m(\n}b_{\rho)} + \f{1}{18}g_{\n\rho}\l( -5 b_\m +\f 7 2 u_\m \r) - \f{5}{18} g_{\m(\n}u_{\rho)} +2 q_{\m(\n\rho)} \ .
\end{align}

\item Using~(\ref{eq:app_conn_decomp})-(\ref
 {eq:disformation}), one finds~\cite{Rigouzzo:2022yan,Rigouzzo:2023sbb} that
 the scalar curvature~(\ref{eq:R_def}) becomes
\begin{align}
\mc R &=  R +  \stackrel{\G}{\nabla_\m} \l( 2 v^\m +u^\m - b^\m \r) - \f 2 3 v_\m \l( v^\m + u^\m - b^\m \r) + \f{1}{24} a_\m a^\m  \nonumber \\
 & \qquad+ \f 1 2 \tau_{\m\n\rho}\tau^{\m\n\rho} -\f{11}{72}u_\m u^\m + \f{1}{18} b_\m b^\m + \f 2 9 u_\m b^\m + \f 1 4 q_{\m\n\rho}\l(q^{\m\n\rho} - 2 q^{\rho\m\n}\r) + \tau_{\m\n\rho} q^{\m\n\rho} \ ,
\end{align} 
with the Ricci scalar given by
\be
R = g^{\sigma\n} \d^\m_\rho \l( \p_\m  \G^\rho_{~\n\s} - \p_\n \G^\rho_{~\m\sigma} + \G^\rho_{~\m\lambda} \G^\lambda_{~\n\sigma} -\G^\rho_{~\n\lambda} \G^\lambda_{~\m\sigma}\r) \ .
\ee

\end{itemize}

\bibliographystyle{utphys}
\bibliography{Refs}

\end{document}